# A Mimetic Strategy to Engage Voluntary Physical Activity In Interactive Entertainment

Andreas Wiratanaya
Gouleystrasse 122b
52146 Würselen
Germany
`wiratanaya@gmail.com`

Michael J. Lyons
College of Image Arts and Sciences
Ritsumeikan University
56-1 Tojiin Kitamachi Kita-ku
Kyoto Japan
`lyons@im.ritsumei.ac.jp`

## Abstract

*We describe the design and implementation of a vision based interactive entertainment system that makes use of both involuntary and voluntary control paradigms. Unintentional input to the system from a potential viewer is used to drive attention-getting output and encourage the transition to voluntary interactive behaviour. The iMime system consists of a character animation engine based on the interaction metaphor of a mime performer that simulates non-verbal communication strategies, without spoken dialogue, to capture and hold the attention of a viewer. The system was developed in the context of a project studying care of dementia sufferers. Care for a dementia sufferer can place unreasonable demands on the time and attentional resources of their caregivers or family members. Our study contributes to the eventual development of a system aimed at providing relief to dementia caregivers, while at the same time serving as a source of pleasant interactive entertainment for viewers. The work reported here is also aimed at a more general study of the design of interactive entertainment systems involving a mixture of voluntary and involuntary control.*

## 1. Introduction

Human-computer interfaces employing face and body gestures as input may be classed into two broad categories: those which take intentional, conciously expressive input from the user and those which process spontaneous, natural actions made without deliberate intention. In the past decade or two, there has been a tremendous interest in the latter class of human-machine interfaces. In particular, the automatic recognition of facial expressions has become a major sub-topic of computer vision and pattern recognition research, as has been discussed in depth in the widely cited reviews of the field [1,2].

The former class of interfaces, designed for voluntary control, have attracted increasing attention more recently, stimulated by the widespread availability of hardware capable of handling real time face and body gesture recognition, as well as efforts to develop methods of human computer interaction which go beyond the keyboard and mouse. To give a specific example of this kind of approach, a series of such works involving intentional action of the face and mouth in voluntary real-time interactive control is summarized by Lyons [3].

While the distinction between voluntary and involuntary interaction may be useful one for system designers, in fact there is no strict dichotomy between these two forms of gestural interaction, but rather a maleable continuum that depends on a user's level of awareness and expertise. Consider, for example, interaces aimed at the processing and interpretation of expressive movement in dance and musical performance, such as is treated in the work of Camurri *et al.* [4], using the EyeWeb platform. It is not difficult to conceive of situations in which a performer or visitor to an installation may be initially unaware that their movements are being tracked and used to control sound synthesis, lighting, or other effects. Perhaps more commonly, a performer may be initially unfamiliar with the properties of a given interactive system and not capable of fully intentional control. With increasing experience and understanding of a system, a user may develop control intimacy [5], whence the mode of interaction will shift from having the character of involuntary recognition of actions, to an expert voluntary 'playing' of that system, just as a skillful musician plays a musical instrument.

In the work reported here, we considered a further possible blending of voluntary and involuntary modes of interaction to design and implement a system which encourages a user to make the transition from a passive, involuntary state, to a physically active, voluntary forms of interaction. In other words, we are interested in developing systems which can capture a subject's attention, engage their interest, then encourage them to play in a voluntary and active fashion. The specific prototype we describe was developed within the context of a larger project aimed at assisting the caregivers and family members of late-stage dementia patients [6, 7]. The general theme of this project is to develop entertainment interfaces suitable for dementia sufferers which will enjoyably hold their interest, thus providing occasional relief for care providers who need periods of time away from fully focussed care in order to attend to other pressing matters.

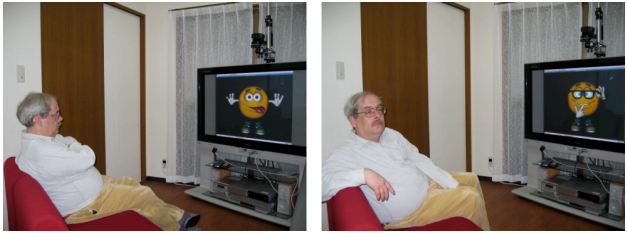

Figure 1: With the iMime system non-verbal interaction is driven by data about a subject's gaze, gestures, and motion qualities as observed using video input from multiple cameras.

While various prostheses exist to help people with physical impairments, the development of technology for impaired cognitive abilities has only recently become the topic of active research [6]. In the case of dementia care the requirement for constant attention can create an unmanageable burden for the patient's family members [7]. The resulting stress can, in turn, have negative effect on the patient's well being. One strategy for reducing the burden of care, as well as the resultant stress, is to keep the patient entertained with audio-visual media that capture and hold the attention for even a limited period of time.

Recently Kuwabara et al. [7] presented a general framework for providing online support for people with dementia or severe memory-impairment giving the concrete application scenario of reminiscence videos which present elderly people with personalized memory stimulating images from their past. To add user interactivity to this application, Utsumi *et al.* [8] explored a content switching strategy for attracting and maintaining the attention of video watchers. This uses the subject's gaze direction as a measure of attention, switching to a different channel whenever the patient starts to lose interest. We have substantially extended this approach by designing and implementing a novel interactive interface with response-dependent adaptation of content display. Instead of reminiscence video contents we are exploring a much more interactive paradigm which uses a real-time lifelike animated character. Within this domain there has been extensive study of embodied conversational agents [9], however there has comparitively little study of purely non-verbal animated characters. Guided by a clinician familiar with the special needs of late-stage dementia sufferers, we have designed a framework for the interaction between a human and a virtual character. Since middle to late stage dementia patients often suffer from severely impaired capacity for verbal communication we base our system primarily on visually-drive non-verbal interaction. This has led us to consider the metaphor of the non-verbal performance of a mime. As can be commonly observed in street performances, mimes are well versed in non-verbal communication. We developed a system, called *iMime*, which draws inspiration from this ability of skilled mime performers to attract and hold attention, and entertain, using non-verbal interactive behaviour. Further support for our approach may be found in the study of Bailenson and Yee [10] in which embodied agents mimicking a viewer's head movements were found to be more persuasive than those which used recorded movements. While the prototype we describe here is still some stages removed from what can be applied in a clinical setting, or the home, valuable lessons have been learned from the design, implementation, and preliminary testing of iMime system.

## 2. System Design

Figure 2 shows a schematic of our iMime system. It illustrates the interaction flow between a human subject and a virtual character generated by a real-time animation engine. Our design is motivated by the performances of street mimes. In street mime a performer often makes a few stylized or humorous actions then freezes. To prompt further motion an observer is expected to engage in some form of active response: either by making a financial donation or reacting to the mime's behaviour. With this simple interaction strategy, a mime is able to bootstrap a completely non-verbal dialogue with strangers without recourse to speech.

We implemented this interaction metaphor as follows: a subject's movement is recorded by multiple video cameras at different scales. Computer vision algorithms are used to analyze the appearance and movement of the subject and draw conclusions about his/her attentional state. This information is passed to a state machine, updated with online reinforcement learning, to determine which behavior the virtual mime should exhibit next. In our prototype a set of animations were designed to be "mimesque", that is, to be entertaining and to encourage the viewer to respond with a reaction that in turn serves as an input to the vision system, hence closing an interaction loop.

### 2.1. Sensory Input

While a variety of sensors can be used to capture information about the attentional state of user, it was decided to restrict input to non-verbal and non-intrusive communication channels. In our prototype we use two video cameras to capture views of the patient at different scales: one camera is focused on the face while the other one captures a view of the entire upper body.

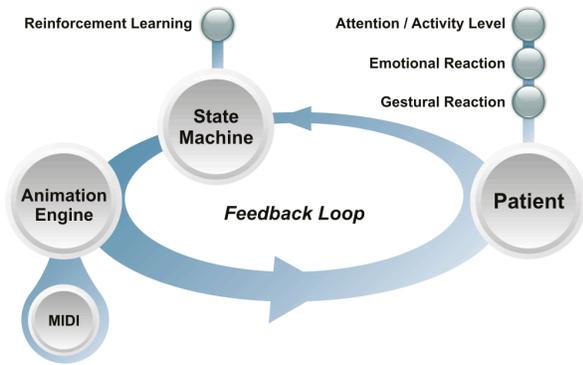

Figure 2: iMime system schematic.

## 2.2. Evaluating Attention

Humans show different signs of attention depending on their level of interest, ranging from simple observation over mild interest up to rapt attention. A subject passively viewing at the screen will merely look at the presented animation, whereas a deeply interested user may show unconscious facial reactions. An actively engaged user may further respond with voluntary body gestures. The vision system implemented for the iMime prototype is capable of: (a) determining whether or not the user is looking at the display, (b) recognizing the subject's current head orientation (c) classifying the subject's overall body motion (d) recognizing emotion primitives such as smile or frown (e) recognizing various basic gestures.

## 2.3. Adapting Character Behaviour to the Viewer

A general limitation of scripted animation systems or video contents is repetitiveness. Even the best scripts become boring after some time if the subject observes a non-changing, repeating pattern. Street mimes observe the reaction of the audience and adapt their behavior appropriately based on experience. It is interesting to notice that a mime will often show an act which by itself is not perceived as being funny or interesting but can be very entertaining in combination with later acts. We simulate this decision process by equipping the state machine that controls the animated character with an online reinforcement learning system. This system analyzes the attentional reaction of the viewer and devises a strategy to to maximize that viewer's attention. Instead of greedily choosing the locally best solution the system is able to select behavioural states which could lead to a better solution in the future even if this involves taking a locally sub-optimal path.

## 2.4. Additional Input Channels

A mime uses different means to entertain her/his audience. While the mime is usually mute during the performance sometimes music is used to augment the gestures and expressions. We included an interface in our iMime prototype that allows for controlling the facial expression of the animated character by playing music files in MIDI format.

## 3. Implementation

### 3.1. Animation Engine

A mime conveys information using the two primary non-verbal channels: facial expression and body language. This has to be kept in mind when choosing a model for the animated character. Through extensive consultation with a clinician involved in research on dementia care, we learned that overly realistic human models would most likely not be acceptable. Animal models, while cute, impose limits on the usable range of body language. Finally, we settled on a relatively abstract model, shown in Figure 3, for its sufficient degree of expressivity, flexibility, and cartoon-like character.

### 3.2. Facial Animation

Facial animation is a widely studied topic in the field of computer graphics and several prior approaches are available. We first implemented a parameterized muscle model. The results were not sufficiently expressive and, in addition, controlling muscle parameters proved to be unintuitive. We therefore decided to use an alternate technique based on handcrafted morph targets. This method decomposes facial expressions into different primitives such as eyebrow raises, mouth shapes, or tongue positions. Figure 4 shows some examples. Blending expression primitives allows for the synthesis of a large variety of composite expressions. The curent model uses 42 different primitives and core expressions.

### 3.3. Body Animation

Limb movements typically involve non-linear rotations. So, the linear morph approach used for facial expression synthesis is not appropriate for synthesizing body poses.

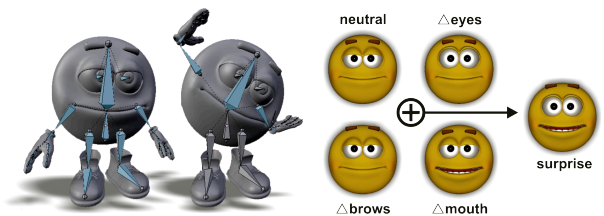

Figure 4: Structural components of the animation engine. Left: the skeletal system used for animating the body. Right: several facial expression morph targets.

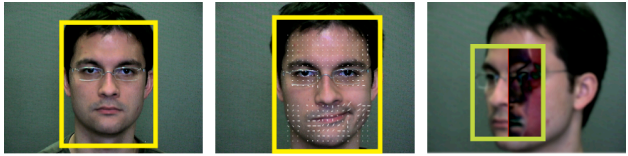

Figure 5: Facial cues used by iMime: Left: face position detection. Center: classifying non-rigid motion using optical flow. Right: attention classification using the symmetric difference image.

Still, we would like to be able to combine movement primitives to generate variety. A widely used solution to this problem is the skeletal animation approach. A skeleton consisting of a hierachical arrangement of bones, inspired by the human body, augments the model. Moving the upper arm then move the lower arm, which in turn moves the hand. Each bone is assigned a region of influence in the model. Figure 4 shows the skeleton used to animated our character. With this approach, Animations can be parameterized and blended easily given the rotation angles for each bone. Adding Perlin noise [11] to each animation parameter, via convolution, gives a more lifelike appearance to the character.

### 3.4. Vision System

Our prototype uses two cameras to capture the patient at different spatial scales. One camera, aimed at the face, captures emotional and attentional information, while the camera focussed on the upper body extracts information about upper body gestures.

### 3.5. Analyzing the Face

We extended a system previously developed by our group [12,13] to analyze information visible in the viewer's face. The system combines automatic face detection and optical flow to classify facial expressions and to discriminate between rigid and non-rigid movement of the head.

For each frame we start by detecting the position of the face using the method of Viola and Jones [14]. We divide the face rectangle into seven regions of interest as shown in Figure 5: eyes, eyebrows, cheeks and mouth. In each region we compute the optical flow over previous frames [12]. Facial movement patterns can then recognized by classifying the characteristic flow vector comprised of the average flow vector for every region. We use patterns of optical flow to discern between non-rigid motion caused by facial expressions and rigid motion caused by head movement [13]. Translation of the head is detected by examining the motion of the face rectangle over the preceding frames. More generally, an analysis of the distribution of the main peaks in the flow field yields a good rigid motion criterion: a wide spatial distribution of motion peaks over the entire face indicates rigid motion. Similarly high average flow in all seven zones indicates presence of overall motion of the face.

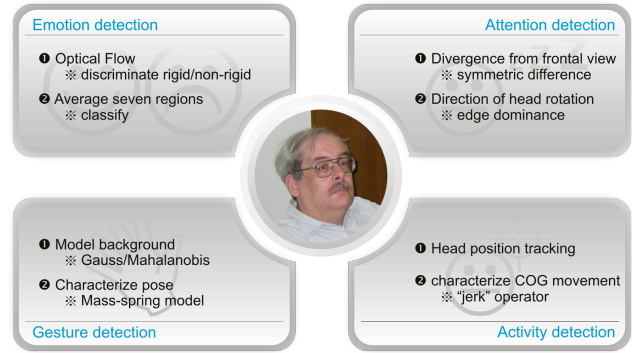

Figure 6: Overview of the vision system

To determine whether the patient is looking at the camera we exploit the symmetry of the human face. The face rectangle returned by the face finder is centered very precisely on the face. A good estimate of the face orientation can be obtained by computing the pixel distance between the left and right halves of the face after applying a median filter to the image. This is illustrated in Figure 5. We also employ an alternative method to determine left/right head orientation. If the head is turned to the left, the face detection rectangle moves so that one side of the rectangle includes the profile of the face, while the other side of the rectangle contains a relatively smooth area. In this case, there is a stronger presence of visual edges in the left side of the rectangle. Computing the center of gravity of all significant edges in the face region therefore provides an indicator for head orientation. Combining these two techniques described above yield a robust estimator of head orientation: while the first algorithm assumes roughly uniform lighting of the face, the second algorithm is insensitive to moderate lighting variations.

Finally to get an estimate of the spatial movement of the viewer we compute the fourth derivative ("jerk" operator) of the head position.

### 3.6. Analyzing the Upper Body

To classify the patient's pose we first separate the foreground pixels (patient) from the background pixels. The envisioned application of the system is in an indoor environment and we use a Gaussian model [15] to describe the background, in which each pixel is described statistically. Figure 7 shows the typical results after removing the background from the input image based on the Mahalanobis distance from a threshold. The algorithm then binarizes the image the current and a mass-spring model [16] is used to recover the characteristic shape of the current pose. This process can be compared to dropping a piece of cloth from the top of the image onto the foreground object. Gravity pulls the cloth down over the object, which holds it in place. A converged drape of the body outline is shown in the right image of Figure 7. The algorithm returns a vector of height values that are normalized and correlated to previously acquired reference poses for classification.

### 3.7. Learning Module

A further novel feature of the platform we studied is that it attempts to learn from interactions with viewers in order to increase viewer attention. Online, real-time re-inforcement learning on the attentional behaviour of the viewer is used to update the transitions between states of the animated character in such a way as to increase it's attractiveness.

We implemented a reinforcement learning system which consisting of two components: a *model estimator* to compute transition probabilities between Markov states based on analysis of human behavior data provided by the vision system, and a *value estimator* which computes the optimal policy based on the current state of the model estimator. The reinforcement-learning problem is equivalent to making optimal decisions in a Markov decision process, in which the world can be summarized in terms of its current state. In each state an agent is allowed to choose an action. Based on the action choice, the agent will enter a new state.

In essence, the learning algorithm rewards the adaptive system for capturing the interest of the subject. A simple operational definition of the viewer's "interest" is the time spent looking at the animated character. More concretely, looking at the animated character yields a reward of 1, whereas looking in a different direction gives a reward of 0. In practice, the reward is calculated at the same time as the adaptive agent is trying to decide which action to take next. This gives the viewer time to react to a change in state of the animated character's behavior, and allows the adaptive system to properly calculate changes based of the viewer's response. The job of the adaptive agent is to increase the reward, and thereby the user's attention.

### 3.8. Model Estimator

Suppose that the interactive character is a given state, for example the action "Routine 1" with the viewing attending, and chooses a certain action, for example switching to "Routine 3". Two outcomes are possible, the viewer will either continue to pay attention to the character or not. We can statistically model the observable outcome as a binomially distributed random variable, and estimate the distribution parameters using Bayes law with uniform priors. The *Model Estimator* of the reinforcement-learning module does exactly these tasks. For each possible state of the human-machine system *(Routine number, Viewer Attending or not Attending),* we need to estimate the parameters that describe this binomial distribution. The model estimator consists of a table of outcomes, representing the statistics for each binomial distribution – namely, the number of times the subject stopped paying attention after a particular change from one animated routine to another, as well as number of times the subject started paying attention to the animated character after such a change. This allows us to estimate the Markov transition probabilities as the maximum a-posteriori parameter values for each binomial distribution.

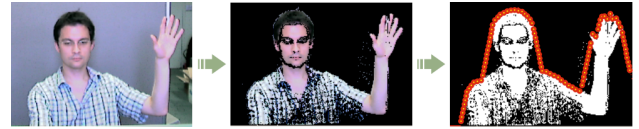

Figure 7: Gesture analysis. Left: input image. Center: background removal. Right: draping the outline.

### 3.9. Value Estimator

The *Model Estimator*, at any given point in time, estimates the transition probabilities for the Markov decision process used by the adaptive animated character when switching behavior routines. Recall that the *Value Estimator* estimates a value function for a given set of model parameters. This value function, once determined, allows us to select the optimal action for the animated character. The popular Q-learning algorithm [17], running in a separate thread, allowed efficient real-time update of the value function. Whenever the model estimator is updated, based on the most recent human-machine interaction, the value estimator is updated to reflect the new model. Thus our adaptive character animation engine always makes optimal decisions based on experience interacting with the viewer.

### 3.10. Action Policy

We programmed iMime to change its action, with a 50% probability, every two seconds. Previous transitions are recorded, as is the viewers state of attention, or inattention. We use this data to update the model estimator module of the learning algorithm. With the model update and value estimator update, iMime looks at the optimal choice and selects it with a probability of $1 - \varepsilon$. It may also select any of the other, non-optimal, choices, with a probability of $\varepsilon$. This procedure is known as the "epsilon-greedy" policy [17]. In this work, a value of $\varepsilon = 0.125$ was used.

## 4. Preliminary Results and Discussion

In order to get a first impression of the possible interactions, we have integrated all components into a prototype system with several video-based interaction modes. The iMime system is aware of user presence. If no user is seen the iMime animated character will either let his gaze wander around the room, showing randomly generated idle body movement or sit down bored and impatiently drum on the floor with his fingers. When a person enters his field of vision he will track him with his eyes and beckon to him. While the user is approaching, the character will indicate by gestures exactly how far he should stand from the camera in order to give a good view of the face. During interaction the character mimics facial expressions such as a smile or eyebrow movement and gestures such as waving. If the user does not interact for a certain period of time the character will first visibly ponder, then point at him and display an example gesture followed by a reward animation if it is mimicked by the viewer. iMime displays a scolding animation if the gesture is not mimicked by the viewer. If the viewers movements are erratic and rapid, iMime stops what he is doing, and gives a puzzled look while scratching his head. These behaviors are intended to be generic and intuitively understandable without verbal explanation.

The state transition machine is designed specifically to enable the transition of passive observation by the subject to active participation in the form of physical gesturing and movement of the subject as s/he attempts to elicit a novel behaviour from the virtual mime. Specifically the system instantiates a simplified non-verbal version of the children's game 'Simon Says': mimicking the gestures of the animated character result in a reward response while the subject is scolded if the response is incorrect.

We have observed the transition from passive to active interaction in informal preliminary tests of the prototype with naïve subjects, however further refinement of the system is necessary before further more definitive tests can be conducted.

## 5. Outlook

This paper described the design and implementation of a character animation system intended to attract, entertain, and engage viewers in a purely non-verbal way. Currently, all core modules are operational and have been integrated into a functional prototype. Preliminary tests with naïve users showed that the system is stable and works as intended. The reinforcement-learning module, however, adapts rather slowly in response to user behaviour. Future work involves investigating strategies for improving the rate of adaptation and, once this is accomplished, conducting further tests with subjects. While the system was designed in the context of a project aimed at dementia care, our more general intention is the development of a system which can serve as an effective testbed for studies of non-verbal interaction in encouraging active volutary physical involvement of users.

## 6. Acknowledgements

We thank Nick Butko for his contributions to the learning module, and Kiyoshi Yasuda for helpful discussions.